\documentclass{elsart3}

\def\a{\alpha}		 		   \def\g{\gamma}	
		 \def\e{\epsilon}                 
		 	   		
		 \def\D{\Delta}	         \def\W{\Omega}	
\def\G{\Gamma}	
                   
         \def\cG{\mathcal{G}}         \def\cF{\mathcal{F}}  
\def\Gc{\mathcal{G}_c}      \def\Gs{\mathcal{G}_s}       
	         	   
       \def\eF{\e_{\rm F}} 	   
\def\kF{k_{\rm F}}	   \def\pF{k_{\rm F}}
\def\[{\left[}           \def\]{\right]}	   
\def\({\left(}           \def\){\right)}	   
\def\<{\langle}          \def\>{\rangle}

\usepackage{epsfig}

\begin{document}

\begin{frontmatter}



\vskip -2.6cm

\title{Quantum interference due to crossed Andreev reflection\\
in a $d$-wave superconductor with two nano-contacts}

\author{S. Takahashi, T. Yamashita, and S. Maekawa}

\address{Institute for Materials Research, Tohoku University, Sendai 980-8577, Japan }


\begin{abstract}
The crossed Andreev reflection in a hybrid nanostructure consisting
of a $d$-wave superconductor and two quantum wires is theoretically
studied.  When the (110) oriented surface of the superconductor is
in contact with the wires parallel and placed close to each other,
the Andreev bound state is formed by the crossed Andreev reflection.
The conductance has two resonance peaks well below the gap structure
in the case of tunnel contacts.  These peaks originate from the bonding
and antibonding Andreev bound states of hole wave functions.
\end{abstract}


\end{frontmatter}

\section{Introduction}

Andreev reflection at the interface of a normal metal and a
superconductor (SC) \cite{andreev,BTK} is one of the fundamental
consequences of superconductivity.  This phenomenon
corresponds to the scattering process of an incoming electron
from the normal side being reflected as a hole,
thereby creating a Cooper pair in the condensate.
In a tunnel junction of a normal metal and a (110) oriented
$d$-wave SC, the zero bias conductance peak appears
due to the formation of the Andreev bound state near the
interface \cite{tanaka}.

When a single quantum wire of a single conducting channel is
in contact with the (110) oriented surface of a $d$-wave SC, the Andreev
reflection is completely suppressed due to the quantum mechanical
diffraction of electron waves at the narrow opening \cite{takagaki}.
A basic question arises what happens if a second quantum wire is 
placed parallel and close to the first one (see Fig.~1).  
For an electron incident from one of the wires into SC, we expect
that the Andreev hole is reflected back into another wire
due to the non-local effect called the {\it crossed} Andreev
reflection (CAR) \cite{bayers,deutcher,falci,melin,yamashita,sanchez}.
Recently, crossed Andreev reflection has been observed in an 
$s$-wave SC with two ferromagnetic wires \cite{beckmann}.

In this paper, we explore the quantum-interference effect due to the
crossed Andreev reflection in a $d$-wave SC with two quantum wires
parallel and close to each other.
It is shown that the resonance peak in the conductance is split into
two sharp peaks at low energies when the barrier potential of the
contact is sufficiently large.  The lower and higher
energy peaks correspond to the bonding and the antibonding Andreev
bound states of quasiparticle holes.

\section{Model and formulation}
\begin{figure}[t]						
  \vskip 0.40cm
  \epsfxsize=0.88\columnwidth				
  \centerline{\hbox{\epsffile{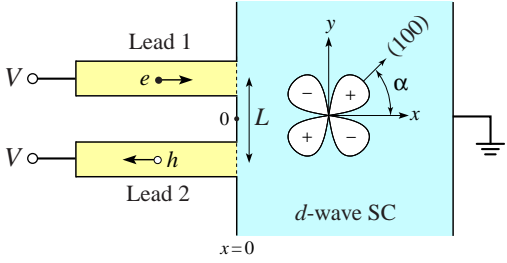}}}	
  \caption{ \small						
Schematic diagram of two normal-conducting quantum wires in	
contact with a $d$-wave superconductor.				
  }   \label{fig1}						
\end{figure}

We examine the quantum transport in a hybrid nanostructure of a
$d$-wave SC and two quantum wires in Fig.~\ref{fig1}.
A two-dimensional (2D) $d$-wave SC occupies the right half space,
and two quantum leads 1 and 2 of width $w$ are connected to SC at
$y=\pm L/2$.
The wave functions of quasiparticles (QP) with excitation energy $E$
in the electrodes are calculated from the Bogoliubov-de Gennes equation.
For simplicity, the Fermi wave number $\kF$ and the effective mass
$m$ are common for all electrodes, and the amplitude of the gap function
is uniform in SC and vanishes in the wires.
Our calculation is restricted to the case where only the lowest subband
is occupied by electrons or holes in the wires.  
When an electron with energy $E$ and wave number
$k_{1} = [2mE+\kF^2 - (\pi/w)^2]^{1/2}$ is incident from lead 1
into SC, the wave functions in leads 1 and 2 are given by
$\Psi_{1}(x,y)=\varphi_{1}(x)\chi(y-L/2)$ and
$\Psi_{2}(x,y)=\varphi_{2}(x)\chi(y+L/2)$ with
\begin{eqnarray}
  \varphi_{1} &=&
      \[ \matrix{1 \cr 0} \] e^{ ik_{1}x}
       + r_{11}^{ee}\[ \matrix{1 \cr 0} \] e^{-ik_{1}x} 
       + r_{11}^{eh}\[ \matrix{0 \cr 1} \] e^{ ik_{1}x}  ,  \cr
  \varphi_{2} &=&
     r_{12}^{ee}\[ \matrix{1 \cr 0} \] e^{-ik_{1}x} 
       + r_{12}^{eh}\[ \matrix{0 \cr 1} \] e^{ ik_{1}x} ,
    \label{eq:phi12}
\end{eqnarray}
and 
$\chi(y)=(2/w)^{1/2} \sin\[(\pi/w)\(y+w/2\)\]$, where
$r_{11}^{ee}$ and $r_{11}^{eh}$ are the amplitudes of the
normal reflection (NR) and the Andreev reflection (AR),
respectively, while $r_{12}^{ee}$ and $r_{12}^{eh}$ are those of
the \textit{crossed} normal reflection (CNR) and the \textit{crossed}
Andreev reflection (CAR), respectively.   
A similar treatment is made for an incident electron from lead 2.
Since $E, \D \ll \eF$, we put $k_{1} \approx \kF [ 1 - (\pi/\kF w)^2]^{1/2}$
in the following.

To carry out the calculation analytically, we make the Andreev
approximation following Blonder-Tinkham-Klapwijk (BTK) \cite{BTK}
to calculate the conductance, and put the wave function of SC in the form
\begin{eqnarray}
  \Psi_{s} =
    \int_{-\kF}^{\kF}
     t_s^{ee}(p_y) \[ \matrix{ 1 \cr  \G_+ } \]
     e^{i\sqrt{\pF^2-p_y^2}x} e^{ip_yy}  dp_y
     \cr   + 
    \int_{-\kF}^{\kF}
     t_s^{eh}(p_y) \[ \matrix{ \G_- \cr 1 } \]
     e^{-i\sqrt{\pF^2-p_y^2}x} e^{ip_yy}  dp_y,
   \label{eq:PsiS}
\end{eqnarray}
where $t_s^{ee}(p_y)$ and $t_s^{eh}(p_y)$ are the transmission coefficients, 
$\G_\pm={\D_\pm / (E+\W_\pm)}$,
$\W_\pm=\sqrt{E^2-|\D_\pm|^2}$,
and $\D_\pm = \D_0\cos2(\theta \mp \a)$
   \cite{tanaka,takagaki},
$\a$ being the angle between the (100) axis of SC and
the normal to the interface (see Fig.~\ref{fig1}), and 
$\theta=\sin^{-1}(p_y/\pF)$ is the propagation angle relative
to the $x$ axis.

In the following, we focus on the (110) oriented surface of SC
($\alpha=\pi/4$) as shown in Fig.~\ref{fig1}.  
In this case, $\D_\pm=\pm\D$ with $\D = \D_0 \sin 2\theta$, and
$\W_\pm=\W=\sqrt{E^2-\D^2}$.
The barrier potential at the interface between the wires and SC
is taken into account by the $\delta$-function-type potential
with amplitude $(\hbar^2\pF/2m)Z$, $Z$ being a dimensionless
parameter of barrier height \cite{BTK}.
The boundary conditions for the wave functions at the interfaces
are $\Psi_{s}(0,y)=\Psi_{i}(0,y)$
and $[\partial_x{\Psi_{s}(x,y)}-\partial_x{\Psi_{i}(x,y)}]_{x=0}=\pF Z\Psi_{i}(0,y)$
 ($i=1,2$) appropriate for the $\delta(x)$ potential.
The matching technique to the boundary conditions \cite{szafer}
yields the reflection coefficients
\begin{eqnarray}
  r_{11}^{ee} &=&  -1  + {\bar{k}_1\over {D}_+}({\bar{k}}_1+\cF-\Gc-iZ)
       \cr \ \ \   & & + {\bar{k}_1\over {D}_-}({\bar{k}}_1+\cF+\Gc-iZ) ,  
    \label{eq:a1}  \\
  r_{11}^{eh} &=& 
       -{\bar{k}}_1{\cG}_s \( {1\over  {D}_-} - {1\over  {D}_+}\),
   \label{eq:b1}  \\
  r_{12}^{ee} &=&
     - {2{\bar{k}}_1\Gc\over  {D}_+ {D}_-} \[ ({\bar{k}}_1+\cF-iZ)^2-(\Gc^2+\Gs^2)\] 
         ,
   \label{eq:a2}  \\
  r_{12}^{eh} &=&
     - {\bar{k}}_1{\cG}_s \( {1\over  {D}_-} + {1\over  {D}_+}\),    
   \label{eq:b2}
\end{eqnarray}
with $\bar{k}_1=k_{1}/\kF$,
${D}_\pm=({\bar{k}}_1+\cF)^2-(\Gc \pm iZ)^2-\Gs^2 $,
\begin{eqnarray}
  \cF &=& 
      \int_{-\kF}^{\kF}   {d{p}\over  2\pi} 
      {\W\over  E}
     \sqrt{1- (p/\kF)^2} \varphi^2(p),
      \label{eq:F1}  \\
  \Gc &=& 
      \int_{-\kF}^{\kF} {d{p}\over  2\pi} 
      {\W\over  E}
     \sqrt{1- (p/\kF)^2} \varphi^2(p)
     \cos(pL) ,
     \label{eq:G1}  \\
  \Gs &=& 
     i \int_{-\kF}^{\kF} {d{p}\over 2\pi} {\D\over E}
     \sqrt{1- (p/\kF)^2} \varphi^2(p)
     \sin(pL) ,
     \label{eq:G2}
\end{eqnarray}
where $\varphi(p)=\<{p}|\chi\>$ is the overlap integral of $\chi(y)$
and $e^{i{p}y}$:
\begin{eqnarray}
  \varphi(p)  = \sqrt{8 w/\pi^2} {\cos(pw/2)/\[ 1-(pw/\pi)^2 \]}.
\end{eqnarray}
Note that $\Gc$ and $\Gs$ represent the non-local
coupling and are dependent on $L$.
In the limit of $L\rightarrow\infty$, where the two contacts are
independent ($\Gc=\Gs=0$), one has
$r_{11}^{ee}=(k_{1}-\cF-iZ)/(k_{1}+\cF+iZ)$, 
$r_{11}^{eh}=r_{12}^{ee}=r_{12}^{eh}=0$,
indicating the complete suppression of AR in a single quantum wire
with a single transverse mode \cite{takagaki}.
The reflection coefficients $r_{22}^{ee}$, $r_{22}^{eh}$, $r_{21}^{ee}$,
and $r_{21}^{eh}$ for an incident electron from lead 2 are 
obtained from those in Eqs.~(\ref{eq:a1})-(\ref{eq:b2}) 
by the replacement $1 \leftrightarrow 2$ and $L \rightarrow -L$.

\vskip -1.0cm
\section{Results and discussion}

\begin{figure}[t]						
  \vskip 0.40cm
  \epsfxsize=0.85\columnwidth				
  \centerline{\hbox{\epsffile{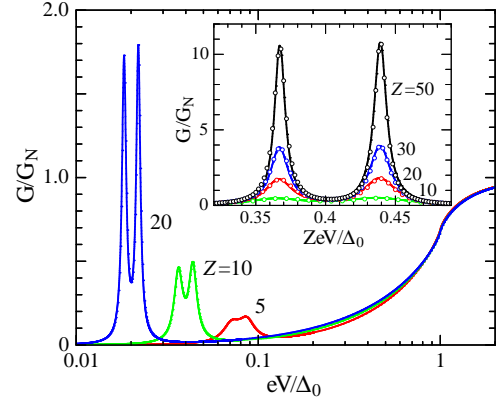}}}	
  \caption{							
Normalized conductance $G/G_N$ as a function of bias voltage $V$
for different values of interfacial barrier parameter $Z$.
Inset shows $G/G_N$ vs normalized voltage $ZV$.
The circles in the inset are calculated from Eq.~(\ref{eq:GsGn}).
  }   \label{fig2}						
\end{figure}

When the bias voltage $V$ is applied to the two leads, 
the conductance $G$ at zero temperature ($T=0$)
is calculated by the formula
\begin{eqnarray}
  G={4e^2\over h} (1-|r_{11}^{ee}|^2-|r_{12}^{ee}|^2
                       +|r_{11}^{eh}|^2+|r_{12}^{eh}|^2),
  \label{eq:Gs}
\end{eqnarray}
where $h$ is the Planck constant and $E=eV$.
Figure~\ref{fig2} shows the conductance $G$ vs $V$ for
$\kF w=4$, $\kF L=8$, and different values of $Z$. \
The conductance is normalized to the normal state value $G_N=G(\D=0)$.
For a low barrier potential of small $Z$, the conductance decreases
monotonically with decreasing $eV$ below $\D_0$.
As the barrier potential becomes higher, a peak structure appears well
below $\D_0$ and shifts towards lower $eV$, developing the double peak
structure with increasing $Z$.
If the conductance is plotted as a function of normalized voltage
$ZV$ as shown in the inset, the conductance peaks fall into the same
position, indicating that the resonance peak positions are
scaled by $1/Z$.

Let us examine the origin of the double peak structure in the conductance
in the tunneling case ($Z \gg 1$).  For $E \ll \D_0$,
$\cF$, $\Gc$, and $\Gs$ in Eqs.~(\ref{eq:F1})-(\ref{eq:G2}) have the forms:
$\cF \approx i f \D_0/E$, $\Gc \approx i g_c\D_0/E$, and $\Gs = i g_s\D_0/E$,
where 
${f} =(E\mathcal{F}/i\D_0)_{E\rightarrow 0}$,
${g_c} =(E\Gc/i\D_0)_{E\rightarrow 0}$, and
${g_s} =(E\Gs/i\D_0)$ are energy independent quantities, 
so that the Andreev reflection coefficients are calculated in the
resonance forms
\begin{eqnarray}  
    r_{11}^{eh} &\approx&
      - { ig\g \over  E-{E_-} + i\g } 
      + { ig\g \over  E-{E_+} + i\g },
  \label{eq:A-b1}  \\
    r_{12}^{eh} &\approx&
      -  { ig\g \over  E-{E_-} + i\g } 
      - { ig\g \over  E-{E_+} + i\g }, 
 \label{eq:A-b2}
\end{eqnarray}
where
$g=g_s/2f$, $E_\pm=  (f \pm |g_c| + g_s^2/2f)\D_0/Z$, and
$\g=(k_{1}/\kF)f\D_0/Z^2$.   
The other coefficients are 
$r_{11}^{ee} \approx -1-gr_{12}^{eh}$ and $r_{12}^{ee} \sim O(g^2)$.
The resonance energies $E_\pm$ and intensity $g$ 
have damped-oscillation dependence on $L$ in the period
of Fermi wave length.  
We note that the resonance positions $E_\pm$ and their separation
$E_+-E_-$ are scaled by $1/Z$, while the line width $\g$
is scaled by $1/Z^2$.   Therefore the line width of the peaks
is much smaller than the separation for $Z \gg 1$, and  a well-separated
two-peak structure is formed as shown in the inset of Fig.~\ref{fig2}.
The conductance $G$ is approximated by the sum of two Lorentzians
\begin{eqnarray}  
   {G \over  G_N}  \approx  
       { {I} \g/\pi \over  (eV-{E_+})^2 + \g^2} 
     + { {I} \g/\pi \over  (eV-{E_-})^2 + \g^2}, 
  \label{eq:GsGn}
\end{eqnarray}
where $I=\pi g^2f\D_0/4\cF_N$ and $\mathcal{F}_N$ is the normal-state value
of $\mathcal{F}$.  The simple formula (\ref{eq:GsGn}) reproduces the numerical
result in the inset of Fig.~\ref{fig2}, if the calculated values
($f=0.405$, $g_c=-0.036$, and $g_s=-0.0345$) are used
in Eq.~(\ref{eq:GsGn}).
Note that the peak height of $G/G_N$ increases in proportion to $Z^2$.

\begin{figure}[b]					
  \vskip 0.40cm
  \epsfxsize=0.96\columnwidth				
  \centerline{\hbox{\epsffile{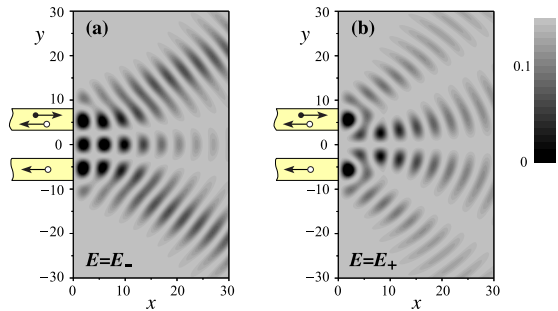}}}	
  \caption{						
Mapping of Andreev bound states on the $xy$ plane,
when an electron is incident from lead 1 to a $d$-wave SC.
(a) and (b) are the absolute square of hole
wave functions, $|\psi^h_{-}|^2$ and $|\psi^h_{+}|^2$,
at resonance energies $E_-$(bonding) and $E_+$(antibonding).
  }   \label{fig3}					
\end{figure}						


To elucidate the formation of the Andreev bound states, we derive an
analytical formula of the QP wave function $\Psi_{s}$ for $Z \gg 1$
and around $E=E_\pm$ in SC:
\begin{eqnarray}
    \Psi_{s}
    \approx  { i\g \over  E-{E_-} + i\g } 
             \[\matrix{\psi_{-}^e \cr \psi_{-}^h}\]
           + { i\g \over  E-{E_+} + i\g } 
             \[\matrix{\psi_{+}^e \cr \psi_{+}^h}\],
   \nonumber
\end{eqnarray}
where, except very close to the interface ($x \approx 0$), 
\begin{eqnarray}
    \[\matrix{\psi_{-}^e  \cr \psi_{-}^h}\] &=& {2g\over \pi}
       \int_{0}^{\kF} {dp} {\D\over E} \varphi(p) \cos{pL\over 2}
      \sin{qx} \[\matrix{\ \ \ \sin py \cr  -\cos py}\],
       \nonumber \\
   \[\matrix{\psi_{+}^e  \cr \psi_{+}^h}\] &=& {2g\over \pi}
       \int_{0}^{\kF} {dp} {\D\over E} \varphi(p) \sin{pL\over 2}
       \sin{qx} \[\matrix{\cos py \cr \sin py}\],
       \nonumber
\end{eqnarray}
with $q=\sqrt{\kF^2-p^2}$ and $\D = 2\D_0(pq/\pF^2)$.
At the lower resonance energy $E_-$, $\Psi_{s}$ is dominated by the first term
whose electron (hole) wave function $\psi^e_{-}$ ($\psi^h_{-}$)
is an odd (even) function of $y$.   
At the higher resonance energy $E_+$, $\Psi_{s}$ is dominated by the second term
whose electron (hole) wave function $\psi^e_{+}$ ($\psi^h_{+}$)
is an even (odd) function of $y$.
These results indicate that the electron (hole) wave functions at the lower
and higher resonance energies have different parity with respect to $y$.

Figure~\ref{fig3} shows the mapping of the absolute squares
$|\psi^h_{\pm}(x,y)|^2$ of QP holes at the resonance energies 
for $\kF w=5$, $\kF L=11$, and $Z=50$.  We see that
the hole wave functions are strongly localized with large peaks
near the contacts due to the formation of the Andreev bound states,
and that the waves are decayed along the (010) and (0${\bar 1}$0)
directions into which a Cooper pair is created.
It is noteworthy that the Andreev holes are reflected back into leads 1
and 2 in phase at $E=E_-$ and out of phase at $E=E_+$
as seen in Eqs.~(\ref{eq:A-b1}) and (\ref{eq:A-b2}).


In summary, we have studied the quantum-interference effect
caused by the crossed Andreev reflection in a hybrid nanostructure 
made up of a $d$-wave SC and two quantum wires.  
When the (110) oriented surface of SC is in contact
with the wires via tunnel barriers, the Andreev bound
states are formed at low energies due to the crossed Andreev reflection,
and two sharp resonance peaks appear in the conductance
well below the superconducting gap structure.
The lower and higher conductance peaks correspond to the bonding
and the antibonding Andreev bound states of the hole wave functions.
The result predicted here is possibly observable in a $d$-wave SC 
with conducting molecular wires, self-organized nanowires, or a STM 
tip with two atomic point contacts.

\bigskip

\noindent{\bf Achnowledgements:}
\medskip

This work was supported by a Grant-in-Aid for Scientific Research
from the Ministry of Education (MEXT), 
the NAREGI Nanoscience project, JSPS, and CREST, Japan.




\end{document}